**Title:**

Electroplating of superconducting films of $MgB_2$ from molten salts

**Authors:**

Hideki Abe,[1] Kenji Yoshii,[2] Kenji Nishida,[1] & Motoharu Imai[1]

[1]*National Institute for Materials Science (NIMS), Tsukuba,*
Ibaraki 305-0047, Japan

[2]*Japan Atomic Energy Research Institute (JAERI),*
*Mikadzuki, Hyogo 679-5148, Japan*

Correspondence should be addressed to H. A.

**Abstract:**

The superconductive boride $MgB_2$ possesses high potential as a practical superconducting material for power-electric products such as superconducing tapes because of its high superconducting transition temperature ($T_c$) of 39 K and its high critical current density ($J_c$) under magnetic fields. Superconducting films of $MgB_2$ have been, however, fabricated only on *planar* substrates with areas of several tens of a square centimeter at the largest so far due to the technical limitation of the conventional fabrication methods. In this report, we describe a technique for the fabrication of $MgB_2$ films by means of electrolysis on molten compounds of Mg and B, that is, *electroplating* of $MgB_2$.

Superconducting films of $MgB_2$ with a $T_c$ of 36 K were successfully electroplated on both the planar and *curved* surfaces using simple installations. The electroplating of $MgB_2$ is expected to open a new route in the field of superconductor engineering because of its ability to fabricate superconducting films on large surfaces of various shapes.

**Text:**

The discovery of superconductivity in $MgB_2$ has attracted extensive interest in the field of superconductor engineering because $MgB_2$ has a very high transition temperature ($T_c$) of 39 K.[1]   In particular, a recent report on the high critical current density ($J_c$) of $MgB_2$ films with an order of $10^5$ $A/cm^2$ at 5 K and 10 T indicated that $MgB_2$ films possess high potential for application to power electrics.[2]   The conventional fabrication methods of $MgB_2$ films are based on vacuum-evaporation, which is one of the best-established methods in the field of semiconductor-processing to fabricate high-quality films under control.[2,3]   The application of vacuum-evaporation has been, however, restricted to planar substrates that are several tens of a square centimeter, at the largest. These restrictions are mainly due to the difficulty with homogeneous evaporation.   A practical technique for the fabrication of superconducting $MgB_2$ films on curved or large-scaled surfaces would be necessary for the production of electric-power products, such as superconducting tapes for high-power magnets[4] or superconducting cavities for high-energy accelerators.[5]

An electrochemical synthetic method for the production of $MgB_2$ has been recently reported.[6-8]   The electrolysis on a fused mixture of $MgCl_2$, $MgB_2O_4$, and KCl enabled the formation of $MgB_2$ by using simple and economical equipment, including a DC power supply and a tubular furnace.[6]   Moreover, the addition of NaCl to the electrolyte resulted in a drastic drop in the residual resistance of the resultant $MgB_2$ bulks.[7]   The $MgB_2$ bulk prepared by the electrolysis has a $T_C$ of 37 K, which is comparable to the intrinsic $T_C$ of $MgB_2$, 39 K.[1]   This suggests that the electrochemical synthetic technique is applicable to the fabrication of $MgB_2$ films with sufficient characteristics required of practical superconducting films, i.e., the *electroplating* of $MgB_2$.  Electroplating is one of the most popular techniques to fabricate conductive films and has been applied to various kinds of materials, ranging from elemental metals, such as copper, to binary semiconductors, such as CdSe.[9,10]  However, as far as is known, there are no reports on electroplating of superconductive compounds.  The process of electroplating is used to fabricate uniform films on large-scale surfaces as well as on irregularly shaped surfaces using cost-effective equipment with high throughput and scalability.  Thus, the application of electroplating with $MgB_2$ has great potential in superconductor engineering.

In this paper, a fabrication technique of $MgB_2$ films by means of electrolysis on molten salts is reported.  Superconducting films of $MgB_2$ with a thickness of several tens of a micrometer have been electroplated successfully onto planar and curved surfaces of graphite substrates.

Figure 1 is a schematic picture of an electrolysis cell.  An alumina crucible with a volume of 30 ml (denoted as A) was used as the electrolyte container.  The crucible was filled with a 20-g electrolyte composed of $MgCl_2$, $MgB_2O_4$, KCl, and NaCl with a molar ratio of 10 : 2 : 5 : 5 (B), according to the optimum composition reported in Ref. (6).  A graphite rod with a diameter of 1.0 mm (C) and a semicylindrical graphite substrate with a diameter of 16 mm (D) were inserted in parallel into the electrolyte powder.  The interval between the axis of the graphite rod C and the flat face of the substrate D was set to be 3.4 mm.  The graphite rod C and the graphite substrate D, which work as the anode and cathode, respectively, were insulated by an insulating ceramic plate (E), in which a ventilation hole (F) was bored.  The graphite anode C and the substrate D were connected to Pt leads with a diameter of 0.1 mm (G).  The electrolysis cell was placed in a tubular furnace (H) with an insertion quartz tube (I) with an inner diameter of 50 mm. One end of the insertion tube (J) was joined to a dry Ar gas line, and the other end was connected to an exhaust (K). The Pt leads were led out of the insertion tube via a hermetic connector to a DC electric power supply outside (L).

In advance of the electrolysis procedure, the electrolyte was thoroughly dried by heating at 400 $^o$C for 1 hour under a dry Ar gas flow. The temperature was then raised to 600 $^o$C, at which the electrolyte powder turned into a colorless transparent fusion. After waiting 30 min for the stabilization of the fusion, the electrolysis process was started by applying a constant DC voltage of 4 V between the electrodes. The amplitude of the electrolysis current decreased gradually from the initial amplitude of the order of 100 mA to the order of 10 mA after 1 hour. During the electrolysis, a colorless irritative gas came out from the surface of the graphite anode C, which was ventilated through the hole F made in the insulating plate E. The electrolysis cell was taken out of the tubular furnace after 1 hour, and the substrate D was removed from the electrolysis cell. Keeping the temperature of the substrate higher than the solidifying temperature of the fused electrolyte, the electrolyte that stuck to the substrate was removed quickly using a centrifuge. After the substrate cooled to room temperature in a dry box, it was immersed in dry DMF (*N*-, *N'*- dimethylformamide) at room temperature for 24 hours, and the residual electrolyte was washed off in dry MeOH (methanol) using an ultrasonic washer.

Figure 2 (a) shows a cross-sectional BSE (back-scattered electron) image of the vicinity of the graphite substrate surface obtained by SEM (scanning electron microscope). The light-gray layer between the graphite substrate and the resin mold corresponds to the electroplated film.    Traces
of the residual electrolyte can be seen as white spots at various places in
Fig. 2 (a).    No crack or exfoliation is recognized at the interface between
the film and substrate, which shows the stiffness of adhesion of the
electroplated film to the substrate.    Figure 2 (b) and (c) show close-up BSE
images of Fig. 2 (a) corresponding to the areas enclosed with white
rectangles denoted by (b) and (c) in Fig. 2 (a), together with the element-
mapping images obtained by electron probe microanalysis (EPMA), which
represent the spatial distribution of the constituent elements, Mg and B.

The BSE and element-mapping images at the area (b) show no apparent
texture (see Fig. 2 (b)), and EPMA indicated that the averaged molar
composition is Mg : B : O = 33 : 32 : 35.    Recent reports on the results of
XPS measurements performed on epitaxially grown films of $MgB_2$ showed
that the surface regions of $MgB_2$ films are partly converted into ionic
compounds of Mg such as $Mg(OH)_2$ as a result of a reaction between the
films and moisture in the air.[11,12]    Thus, it is assumed that a considerable
amount of O detected by EPMA comes from the inclusion of $Mg(OH)_2$ in
the surface region of the film, which originates from a reaction between the
film and moisture in the air or residual water in the solvents to remove the
electrolyte during the preparation of the SEM specimen.    Subtracting the
contribution of $Mg(OH)_2$ from the EPMA data as an inclusion, the
composition of the electroplated film at the area (b) in Fig. 1 (a) becomes
Mg : B = (33-35/2) : 32 = 33 : 67, which is very close to the nominal
composition of $MgB_2$, Mg : B = 1 : 2.    This suggests that the film in area
(b) is intrinsically composed of $MgB_2$.

On the other hand, two phases with different contrasts can be observed
to coexist in area (c) (see the BSE image of Fig. 2 (c)).    One phase with a darker contrast has an island-like shape and is surrounded by the other phase with a brighter contrast.   The intrinsic molar composition of the surrounding phase was estimated to be the same as that of area (b), i.e., Mg : B = 33 : 67.   The island-shaped phase is observed as dark and bright spots in the Mg- and B-mapping images, respectively (see the Mg- and B-mapping images of Fig. 2 (c)).   This shows that the island-shaped phase contains the lesser molar fraction of Mg and the larger molar fraction of B compared to the surrounding phase.   It is assumed that the B-rich inclusion arises from the change in the composition of the fused electrolyte in the vicinity of the substrate surface during electrolysis and, thus, could be suppressed by homogenization of the electrolyte by stirring or bubbling.

The magnetization of the same specimen for Fig. 2 under a field of 0.001 T is shown as a function of temperature in Fig. 3.   The magnetic field was applied parallel to the surface of the film to avoid the shape effect. The dimension of the electroplated film was 1.0 x 1.0 mm in length and width, respectively, and the averaged thickness of the film was estimated to be $40\pm5$ μm by dividing the total area of the cross-section of the film by the length of the film-substrate interface (see Fig. 2 (a)).   The magnetization of the film was evaluated on the basis of the estimation of the dimension of the film, as shown in Fig. 3.   The zero-field-cooled magnetization ($M_{ZFC}$ ) at 5 K is negative, and its amplitude is $-0.72\pm0.06$ emu/cm$^3$.   With increasing the temperature, $M_{ZFC}$ increases gradually, and it becomes nearly equal to zero at 36 K (see the inset).   $M_{ZFC}$ remains at zero and does not show apparent temperature dependence from 36 K up to 60 K. Decreasing the temperature from 60 K under a field of 0.001 T, the field-cooled magnetization ($M_{FC}$) follows the $M_{ZFC}$ curve down to 36 K.   $M_{FC}$ begins to drop to negative at 36 K and continues to decrease with lowering the temperature to 5 K, at which $M_{FC}$ is about 10 % of $M_{ZFC}$.  Taking into account the compositional evaluation by EPMA, it is reasonable to regard the diamagnetism observed in both $M_{ZFC}$ and $M_{FC}$ below 36 K as a consequence of the Meissner effect of the $MgB_2$ included in the film.  The volume fraction of $MgB_2$ in the film is estimated to be $91\pm7$ % from $M_{ZFC}$ of $-0.72\pm0.06$ emu/cm$^3$ at 0.001 T and 5 K, which suggests strongly that the electroplated film is composed mostly of $MgB_2$ except for the surface region in direct contact with the surroundings.

Figure 4 shows the temperature dependence of the electrical resistance R.  With decreasing the temperature from the room temperature, R increases slightly to 6.0 m$\Omega$, and, at 36 K, it begins to drop (see inset).  R decreases monotonously down to 31 K, below which it becomes negligibly small.  Thus, the magnetization and resistance measurements showed that the $T_c$ of the electroplated $MgB_2$ films is 36 K.  The slightly low $T_c$ and broad superconducting transitions of the electroplated $MgB_2$ films are thought to be due to the inclusion of O and the B-rich phase (see Fig. 2 (c)) in the surface regions of the films.

Figure 5 gives a typical example of the application of electroplating from the molten salts to a curved surface.  A coaxial electrode unit composed of a graphite anode with a diameter of 1.0 mm and a graphite tubular cathode with an inner diameter of 3.0 mm (see the schematic picture in Fig. 5) was immersed into the molten electrolyte that was electrolyzed in the same condition as the electroplating on the planar surface shown in Fig. 2.  Magnetization measurements have confirmed that the film electroplated on the inner wall of the cathode with a thickness of 40 $\mu$m was mostly composed of $MgB_2$ with a volume fraction higher than 90 %, which shows the ability of electroplating from the molten salts to apply $MgB_2$ films on the curved as well as the planar surfaces.

In summary, superconducting films with $T_c$ of 36 K were electroplated onto the planar and curved surfaces of graphite substrates by electrolysis on a fused mixture of $MgB_2O_4$, KCl, NaCl, and $MgCl_2$ at 600 $^o$C in an Ar atmosphere.   EPMA and magnetization measurements indicated that the films consist mostly of $MgB_2$, except for the surface regions that contain O and the B-rich phase as inclusions.   The electroplating of $MgB_2$ is easy to scale-up because of the simple and low-priced installations, which suggests that the method is applicable to the industrial fabrication of $MgB_2$ films on large-scale surfaces of various shapes.

**Methods:**

The microscopic analysis on the electroplated films was carried out using an EPMA apparatus (JXA-8900R, JEOL Co., Ltd.).   The magnetization measurements were performed using a superconducting quantum interference device (SQUID) magnetometer (MPMS, Quantum Design Co., Ltd.).   The resistance measurements were performed by a conventional DC four-probe method.   Gold electric leads were set to the surface of the film using conductive silver-paste.

**Acknowledgments:**

This research was supported by the Arai Science and Technology Foundation.

**Figure legends:**

Fig. 1.

(a): Cross-sectional images of the film electroplated on the graphite substrate.

(b, c): Close-up BSE images of the parts enclosed with rectangles in Fig. 2 (a) together with the element-mapping images of Mg and B.

Fig. 2.

Temperature dependence of magnetization of the film electroplated on the graphite substrate at a field of 0.001 T parallel to the film surface.   The solid and open circles represent $M_{ZFC}$ and $M_{FC}$, respectively.   The inset shows a magnified view of $M_{ZFC}$ at around $T_c = 36$ K.   The solid curves are guides for the eye.

Fig. 3.

Temperature dependence of electrical resistance of the film electroplated on the graphite substrate.   The inset shows a magnified view at around $T_c = 36$ K.   The solid curves are guides for the eye.

Fig. 4.

Cross-sectional optical microscope image of the film electroplated on the curved surface of the graphite substrate, together with a schematic picture of the electrode unit (refer to the text).

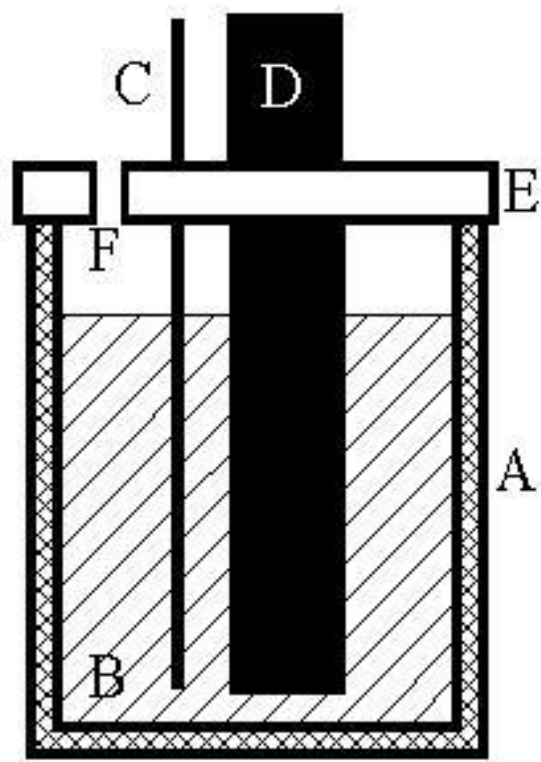

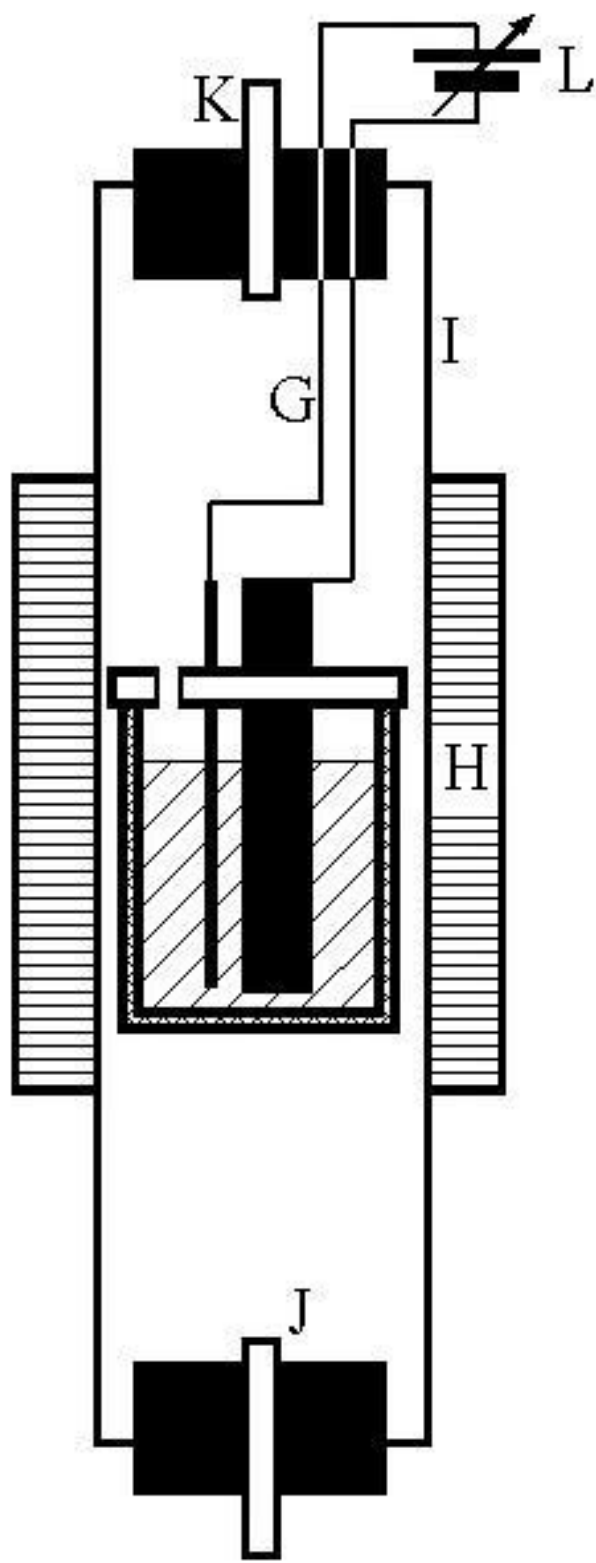

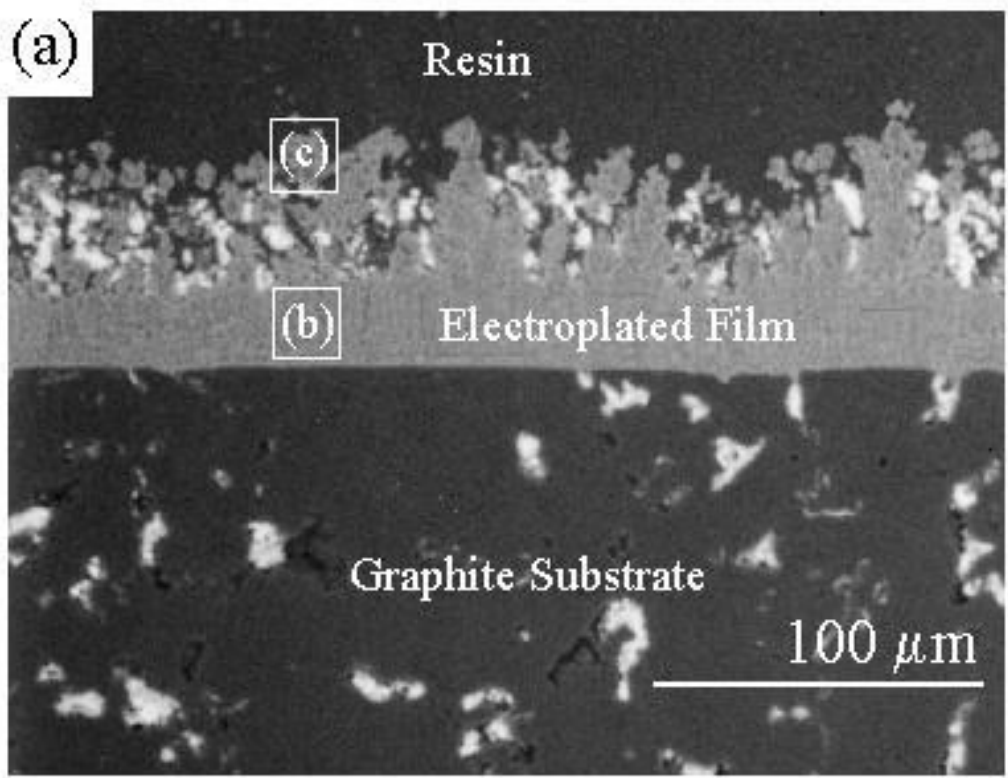

(a)

Resin

(c)

(b)

Electroplated Film

Graphite Substrate

100 μm

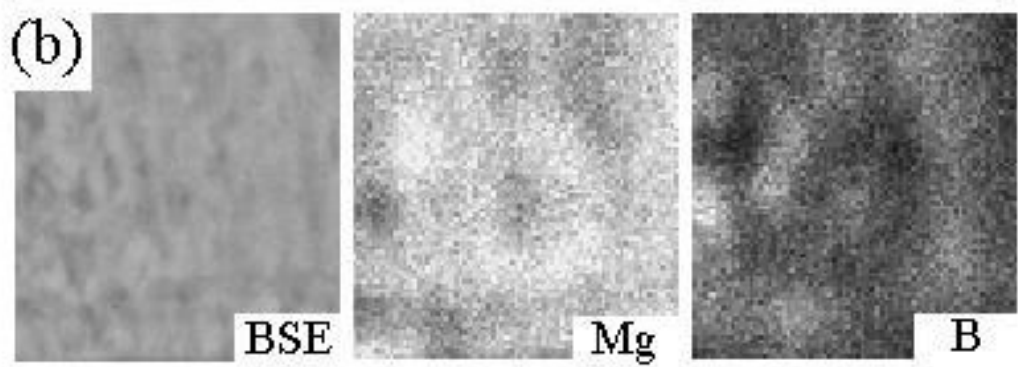

(b)

BSE

Mg

B

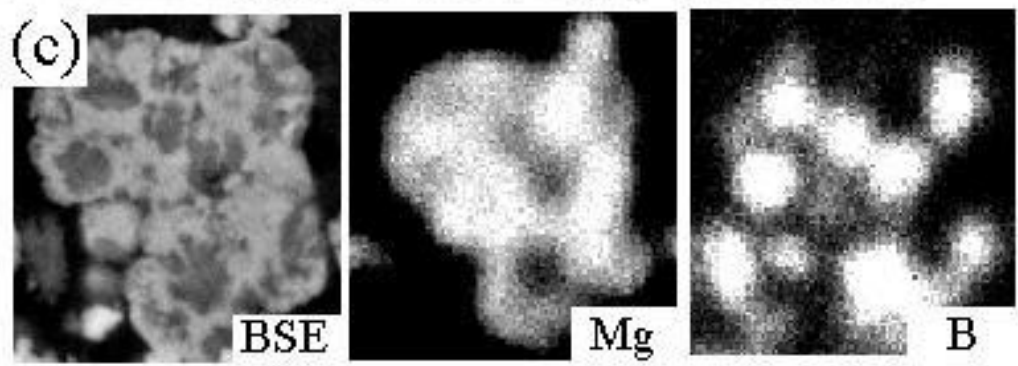

(c)

BSE

Mg

B

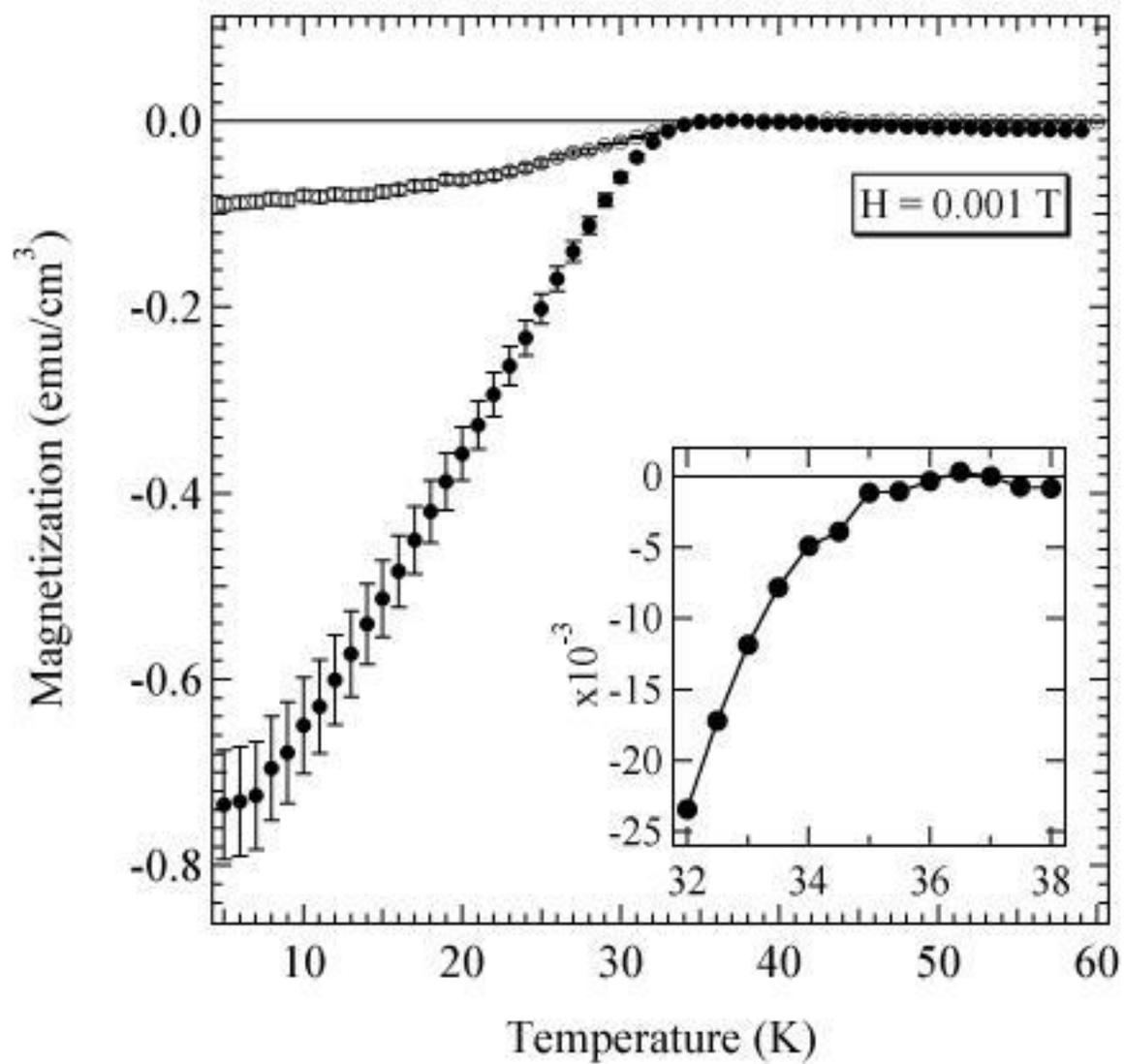

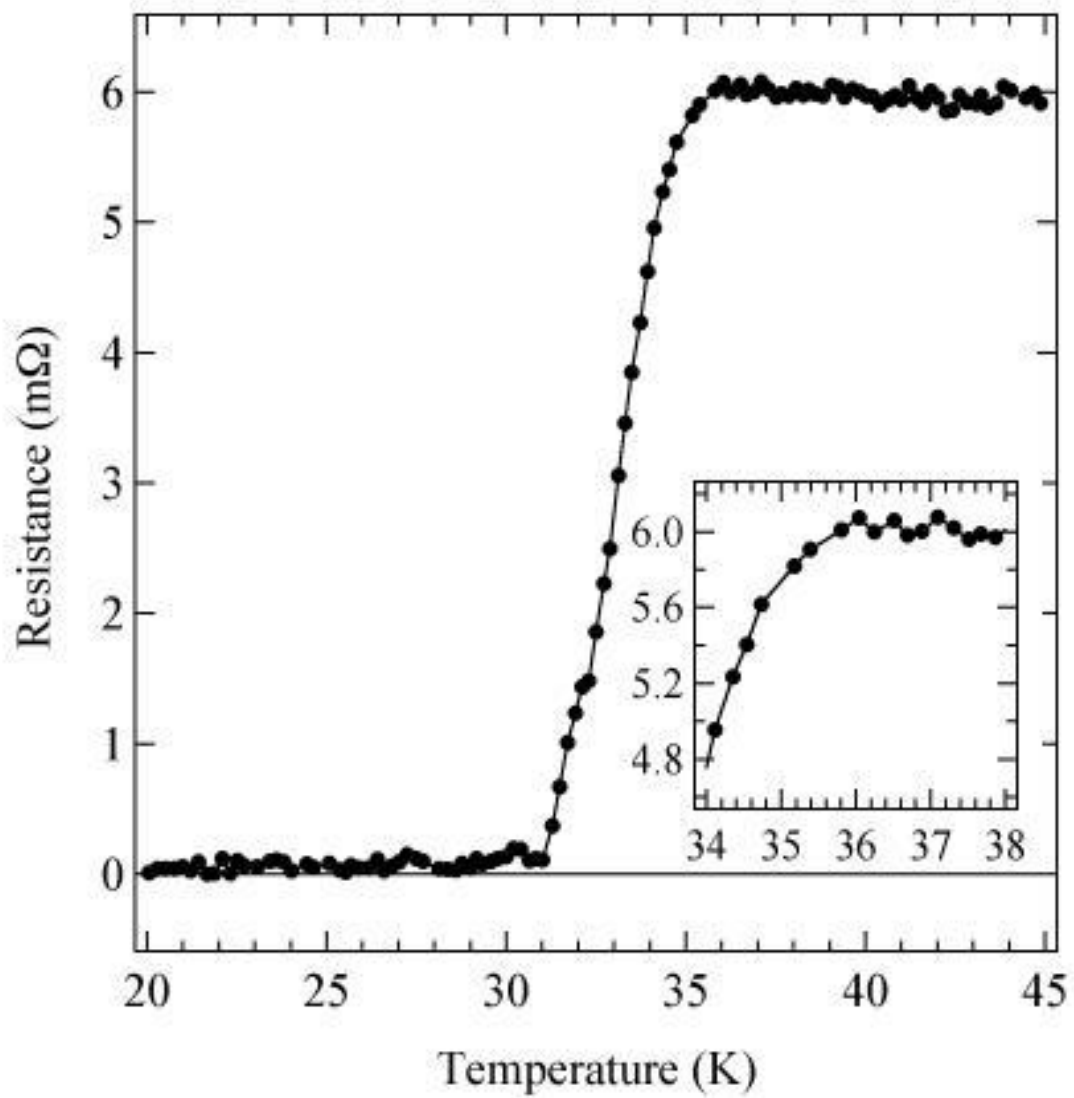

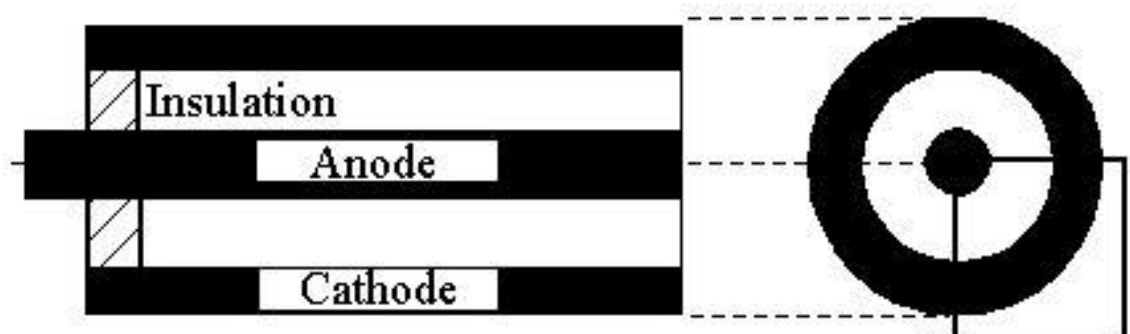

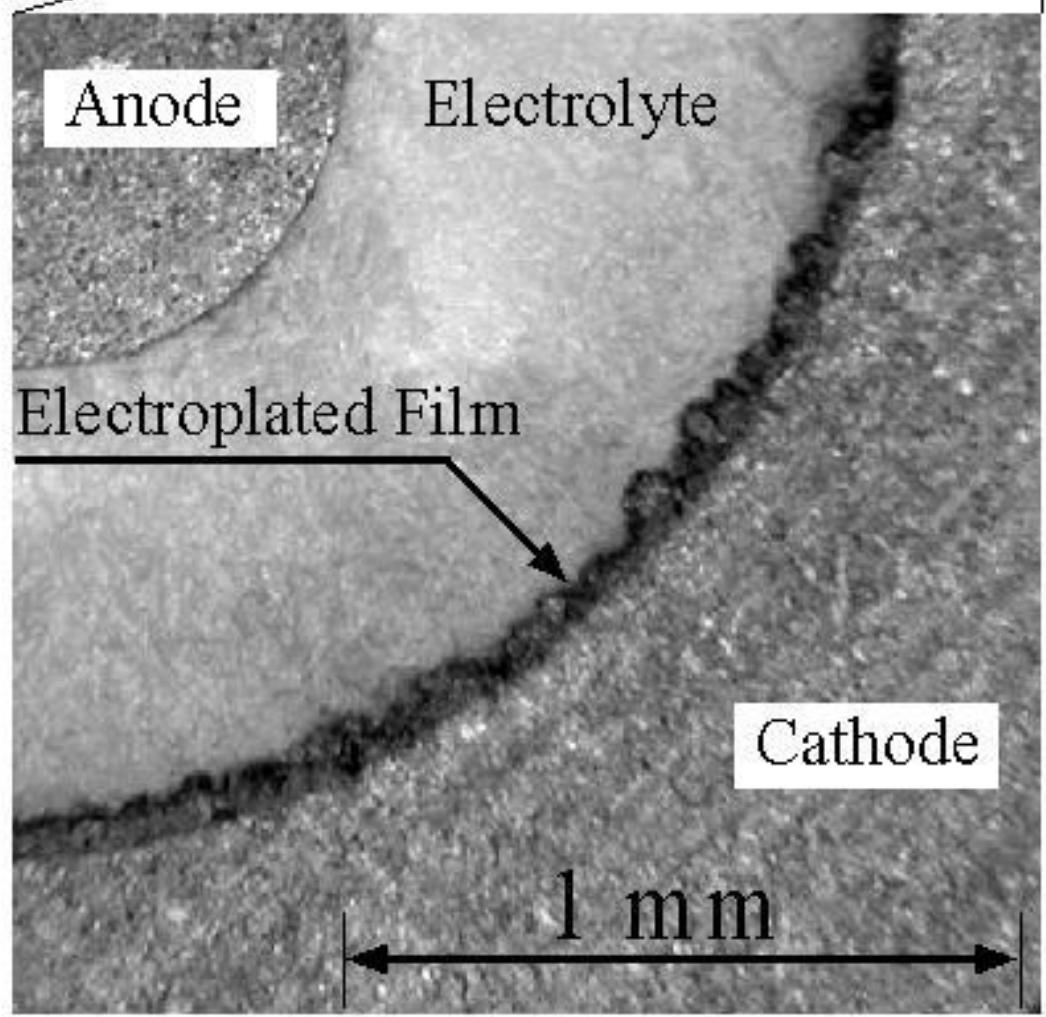

Insulation

Anode

Cathode

Anode

Electrolyte

Electroplated Film

Cathode

1 mm